\documentclass[twocolumn,twoside]{IEEEtran}
\usepackage{amsmath,amssymb,amsthm,wasysym,epsfig,color,subfigure,graphicx}
\usepackage{enumerate,url,algpseudocode,algorithm,balance,url,bm,nicefrac,caption,subfloat}
\usepackage[table]{xcolor}
\usepackage{stfloats}
\usepackage{graphicx}
\usepackage{subfigure}

\theoremstyle{remark}
\allowdisplaybreaks

\newcommand{\minimize}{{\rm minimize}}

\begin{document}
	\title{Two-Timescale Voltage Control in Distribution Grids Using Deep Reinforcement Learning}


\author{Qiuling Yang,~Gang Wang,~\IEEEmembership{Member,~IEEE},~Alireza Sadeghi,\\
	Georgios B. Giannakis,~\IEEEmembership{Fellow,~IEEE}, and Jian Sun,  \IEEEmembership{Member,~IEEE}
	
	\thanks{Manuscript received April 19, 2019; revised May 28, and August 29, 2019; accepted October 31, 2019. The work of Q. Yang and J. Sun was supported in part by the National Natural Science Foundation of China under 
		Grants 61522303, 61720106011, and 61621063.
		Q. Yang was also supported by the China Scholarship Council. The work of G. Wang, A. Sadeghi, and G. B. Giannakis was supported by  
		National Science Foundation 
		under Grants 1509040, 1711471, and 1901134. 
		
		Q. Yang and J. Sun are with the State Key Lab of Intelligent Control and Decision of Complex Systems, 
		School of Automation, 
		Beijing Institute of Technology, Beijing 100081, China (e-mail: yang6726@umn.edu, sunjian@bit.edu.cn). 	
		G. Wang, A. Sadeghi, and G. B. Giannakis are with the Department of Electrical and Computer Engineering, University of Minnesota, Minneapolis, MN 55455, USA
		(e-mail:  gangwang@umn.edu, sadeghi@umn.edu, georgios@umn.edu).
	}
}

\markboth{IEEE TRANSACTIONS ON SMART GRID (Accepted, October 31, 2019)}{}
\maketitle

\begin{abstract}
	Modern distribution grids are currently being challenged by frequent and sizable voltage fluctuations, due mainly to the increasing deployment of electric vehicles and renewable generators. Existing approaches to maintaining bus voltage magnitudes within the desired region can cope with either traditional utility-owned devices (e.g., shunt capacitors), or contemporary smart inverters that come with distributed generation units (e.g., photovoltaic plants). 
	The discrete on-off commitment of capacitor units is often configured on an hourly or daily basis, yet smart inverters can be controlled within milliseconds, thus challenging joint control of these two types of assets. In this context, a novel two-timescale voltage regulation scheme is developed for distribution grids by judiciously coupling data-driven with physics-based optimization. On a faster timescale, say every second, the optimal setpoints of smart inverters are obtained by minimizing instantaneous bus voltage deviations from their nominal values, based on either the exact alternating current power flow model or a linear approximant of it; whereas, on the slower timescale (e.g., every hour), shunt capacitors are configured to minimize the long-term discounted voltage deviations using a deep reinforcement learning algorithm. Extensive numerical tests on a real-world $47$-bus distribution network as well as the IEEE $123$-bus test feeder using real data corroborate the effectiveness of the novel scheme. 
\end{abstract}


\section{Introduction} \label{Sec:Intro}

Frequent and sizable voltage fluctuations caused by the growing deployment of electric vehicles, demand response programs, and renewable energy sources, challenge modern distribution grids. Electric utilities are currently experiencing major issues related to the unprecedented levels of load peaks as well as renewable penetration. For instance, a solar farm connected at the end of a long distribution feeder in a rural area can cause voltage excursions along the feeder, while the apparent power capability of a substation transformer is strained by frequent reverse power flows. Moreover, over-voltage happens during midday when photovoltaic (PV) generation peaks and load demand is relatively low; whereas voltage sags occur mostly overnight due to low PV generation even when load demand is high~\cite{carvalho2008distributed}. This motivates why voltage regulation, the task of maintaining bus voltage magnitudes within desirable ranges, is critical in modern distribution grids.

Early approaches to regulating the voltages at a residential level have mainly relied on utility-owned devices, including load-tap-changing transformers, voltage regulators, and capacitor banks, to name a few. 
They offer a convenient means of controlling reactive power, through which the voltage profile at their terminal buses as well as at other buses can be regulated~\cite[p. 678]{kundur1994power}. 
Obtaining the optimal configuration for these devices entails solving mixed-integer programs, which are NP-hard in general. To optimize the tap positions, a semi-definite relaxation heuristic was used in~\cite{robbins2016optimal,bazrafshan2018optimal}. 
Control rules based on heuristics were developed 
in \cite{tziouvaras2000mathematical,carvalho2008distributed}. 
However, these approaches can be computationally demanding, and do not guarantee optimal performance. A batch reinforcement learning (RL) scheme based on linear function approximation was lately advocated in~\cite{xu2018optimal}. 


Another characteristic inherent to utility-owned equipment is their limited life cycle, which prompts control on a daily or even monthly basis.
Such configurations have been effective in traditional distribution grids without (or with low) renewable generation, and with slowly varying load. Yet, as distributed generation grows in residential networks nowadays \cite{su2014stochastic},  \cite{ipakchi2009grid}, 
rapid voltage fluctuations occur frequently. 
According to a recent landmark bill, California mandated $50\%$ of its electricity to be powered by renewable resources by $2025$ and $60\%$ by $2030$.
The power generated by a solar panel can vary by 15\% of its nameplate rating within one-minute intervals~\cite{wang2016ergodic}. Voltage control would entail more frequent switching actions, and further installation of control devices. 

Smart power inverters on the other hand, come with contemporary distributed generation units, such as PV panels, and wind turbines. Embedded with computing and communication units, these can be commanded to adjust reactive power output within seconds, and in a continuously-valued fashion. Indeed, engaging smart inverters in reactive power control
has recently emerged as a promising solution \cite{kekatos2015stochastic}.
Computing the optimal setpoints for inverters' reactive power output is an instance of the optimal power flow task, which is non-convex~\cite{farivar2011inverter}.  
To deal with the renewable uncertainty as well as other communication issues (e.g., delay and packet loss),  
stochastic, online, decentralized, and localized reactive control schemes have been advocated~\cite{kekatos2015stochastic,zhu2016fast,kekatos2016voltage,wang2016ergodic,fitee2019wgcs,lin2018real,zhang2018distributed}.   

RL refers to a collection of tools for solving Markovian decision processes (MDPs),
especially when the underlying transition mechanism is unknown~\cite{RLbook}.
In settings involving high-dimensional, continuous action and/or state spaces however, it is well known that conventional RL approaches suffer from the so-called `curse of dimensionality,' which limits their impact in practice~\cite{mnih2015human}.  
Deep neural networks (DNNs) can address the curse of dimensionality in the high-dimensional and continuous state space by providing compact low-dimensional representations of high-dimensional inputs \cite{goodfellow2016deep}. Wedding deep learning with RL (using a DNN to approximate the action-value function), deep (D) RL has offered artificial agents with human-level performance across diverse application domains~\cite{mnih2015human,sadeghi2019optimal}. 
(D)RL algorithms have also shown great potential in several challenging power systems control and monitoring tasks~\cite{diao2019autonomous,ernst2004power,xu2018optimal,zamzam2019energy,yan2018data, lu2019demand}, and load control \cite{tsg2017cvr,duan2018qlearning}.
A batch RL scheme using linear function approximation was developed for voltage regulation in distribution systems~\cite{xu2018optimal}. For voltage control of transmission networks, DRL was recently investigated to adjust generator voltage setpoints \cite{diao2019autonomous}.	
A shortcoming of the mentioned (D)RL voltage control schemes is their inability to cope with the curse of dimensionality in action space.  Moreover, \textit{joint control} of both utility-owned devices and emerging power inverters has not been fully investigated. In addition, the discrete variables describing the on-off operation of capacitors and slow timescale associated with changing capacitor statuses, compared with those of fast-responding inverters further challenges voltage regulation. As a consequence, current capacitor decisions have a long-standing influence on future inverter setpoints. The other way around, current inverter setpoints also affect future commitment of capacitors through the aggregate cost. Indeed, this two-way long-term interaction is difficult to model and cope with.

In this context, voltage control is dealt with in the present paper using shunt capacitors and smart inverters. Preliminary results were presented in \cite{yang2019smartgridcomm}.
A novel two-timescale solution combining first principles based on physical models and data-driven advances is put forth. On the slow timescale (e.g., hourly or daily basis), the optimal configuration (corresponding to the discrete on-off commitment) of capacitors is formulated as a Markov decision process, by carefully defining state, action, and cost according to the available control variables in the grid. The solution of this MDP is approached by means of a DRL algorithm. This framework leverages the merits of the so-termed \textit{target network} and \textit{experience replay}, which can remove the correlation among the sequence of observations, to make the DRL stable and tractable. On the other hand, the setpoints of the inverters' reactive power output, are computed by minimizing the instantaneous voltage deviation using the exact or approximate grid models on the fast timescale (e.g., every few seconds).


Compared with past works, our contributions can be summarized as follows.
\begin{itemize}
	\item[\bf c1)] \textit{Joint control of two types of assets.} A hybrid data- and physics-driven approach to  managing both utility-owned equipment as well as smart inverters;
	\item[\bf c2)] \textit{Slow-timescale learning.} Modeling demand and generation as Markovian processes, optimal capacitor settings are learned from data using DRL; 
	\item[\bf c3)] \textit{Fast-timescale optimization.} Using exact or approximate grid models, the optimal setpoints for inverters are found relying on the most recent slow-timescale solution; and,
	{  \item[\bf c4)] \textit{Curse of dimensionality in action space.} Introducing hyper deep $Q$-network to handle the curse of dimensionality emerging due to large number of capacitors. }
\end{itemize}

%

\section{Voltage Control in Two Timescales}\label{sec:problem}

In this section, we describe the system model, and formulate the two-timescale voltage regulation problem. 

\subsection{System model}\label{subsec:syst}
Consider a distribution grid of $N+1$ buses rooted at the substation bus indexed by $i = 0$,
whose buses are collected into $\mathcal{N}_0:=\{0\} \cup \mathcal{N}$, and lines into $\mathcal{L}:=\{1,\ldots,N\}$. 
{ For all $i\in\mathcal{N}$ (i.e., without substation bus), let $v_{i}$ denote their squared voltage magnitude, and $p_i + j q_i$ their complex power injected. 
	For brevity, collect all nodal quantities into column vectors $\pmb{v}$, $\pmb{p}$, $\pmb{q}$. 
	Active power injection is split into its generation $p_{i}^g$ and consumption $p_{i}^c$ as $p_{i}:=p_{i}^g-p_{i}^c$; likewise, reactive power injection is $q_{i}:=q_{i}^g-q_{i}^c$. In distribution grids, it holds that $p_{i}^g = p_{i}^c = q_{i}^c = 0$ and $q_{i}^g >0$ 
	if bus $i$ has a capacitor; while $p_{i}^g = q_{i}^g =0$ if bus $i$ is a purely load bus; and $p_{i}^c \geq 0$, $q_{i}^c \geq 0$, $p_{i}^g \geq 0$ if bus $i$ is equipped with a DG. Let us stack generation and consumption components into vectors $\pmb{p}^g$, $\pmb{q}^g$, $\pmb{p}^c$, and $\pmb{q}^c$ accordingly.
	Predictions of active power consumption and solar generation $(\pmb{p}^c,\pmb{q}^c,\pmb{p}^g)$ can be obtained through the hourly and real-time market (see e.g., \cite{kekatos2015stochastic}), or by running load demand (solar generation) prediction algorithms \cite{tsp2019psse}.}

\begin{figure}
	\centering
	\includegraphics[width =0.5 \textwidth]{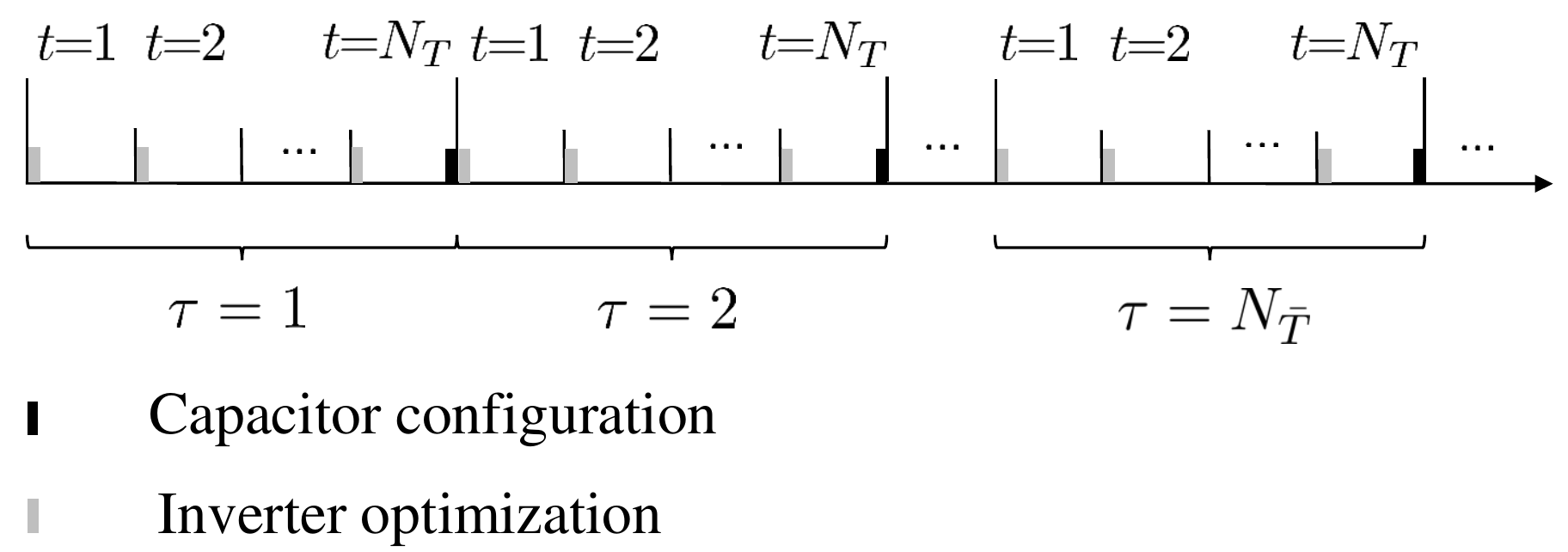}
	\caption{Two-timescale partitioning of a day for joint capacitor and inverter control.}
	\label{fig:twotimescale}
\end{figure}

As mentioned earlier, there are two types of assets in modern distribution grids that can be engaged in reactive power control; that is, utility-owned equipment featuring discrete actions and limited lifespan, as well as smart inverters controllable within seconds and in a continuously-valued fashion. As the aggregate load varies in a relatively slow way, traditional devices have been sufficient for providing voltage support; while fast-responding solutions using inverters become indispensable with the increase of uncertain renewable penetration. 
In this context, the present work focuses on voltage regulation by capitalizing on the reactive control capabilities of both capacitors and inverters, while our framework can also account for other reactive power control devices. To this end, we divide every day into $N_{\bar T}$ intervals indexed by $\tau=1,\ldots,N_{\bar T}$. Each of these $N_{\bar T}$ intervals is further partitioned into $N_T$ time slots which are indexed by $t=1,\ldots,N_T$, as illustrated in Fig.~\ref{fig:twotimescale}. To match the slow load variations, the on-off decisions of capacitors are made (at the end of) every interval $\tau$, which can be chosen
to be e.g., an hour; yet, to accommodate the rapidly changing renewable generation, the inverter output is adjusted (at the beginning of) every slot $t$, taken
to be e.g., a minute. 
We assume that quantities 
$\pmb{p}^g(\tau,t)$, 
$\pmb{p}^c(\tau,t)$, and $\pmb{q}^c(\tau,t)$ remain the same within each $t$-slot, but may change from slot $t$ to $t+1$. 

Suppose there are $N_a$ shunt capacitors installed in the grid, whose bus indices are collected in $\mathcal{N}_a$, and are in one-to-one correspondence with entries of $\mathcal{K}:=\{1,\ldots,N_a \}$ (a simple renumbering). Assume that every bus is equipped with either a shunt capacitor or a smart inverter, but not both. The remaining buses, after removing entries in $\mathcal{N}_a$ from $\mathcal{N}$, collected in $\mathcal{N}_r$, are assumed equipped with inverters. This assumption is made without loss of generality as one can simply set the upper and lower bounds on the reactive output to zero at buses having no inverters installed. 

As capacitor configuration is performed on a slow timescale (every  $\tau$), the reactive compensation $q_{i}^g(\tau,t)$ provided by capacitor $k_i\in\mathcal{K}$ (i.e., capacitor at bus $i$) is represented by
\begin{equation}
q_{i}^{g}(\tau,t) = \hat y_{k_i}(\tau) q_{a,k_i}^{g}, \quad \forall i \in \mathcal N_a, \tau,t
\end{equation}
where $\hat y_{k_i}(\tau)\in \{0,1\}$ is the on-off commitment of capacitor $k_i$ for the entire interval $\tau$. Clearly, if $\hat y_{k_i}(\tau)=1$, a constant amount (nameplate value) of reactive power $q_{a,k_i}^{g}$ is injected in the grid during this interval, and $0$ otherwise. For convenience, the on-off decisions of capacitor units at interval $\tau$ are collected in a column vector $\hat {\pmb y}(\tau)$.

On the other hand, the reactive power $q_{r,i}^{g}(\tau,t)$ generated by inverter $i$ is adjusted on the fast timescale (every $t$), and it is constrained by
$|q_{r,i}^{g}(\tau,t)|\leq \sqrt{(\bar s_i)^2-( p^g_i(\tau, t))^2}$, where $\bar s_i$ is the power capability of inverter $i$. Traditionally, inverter $i $ is designed as $ \bar s_i = \bar p^g_i$, where $\bar p^g_i$ is the active power capacity of the renewable generation unit installed at bus $i$. However, when maximum output is reached, i.e., $p^g_i(\tau, t) = \bar p^g_i$, no reactive power can be provided. To address this, oversized inverters' nameplate capacity has been advocated such that $ \bar s_i > \bar p^g_i$ \cite{kekatos2015stochastic}.
For instance, choosing $ \bar s_i = 1.08 \bar p^g_i$ and limiting $q_{r,i}^{g}(\tau,t)$ to $ \sqrt{(\bar s_i)^2-( \bar p^g_i)^2}$ instead of $\sqrt{(\bar s_i)^2-( p^g_i(\tau, t))^2}$, the reactive power compensation provided by inverter $i$ is $|q_{r,i}^{g}(\tau,t)|\leq 0.4 \bar p^g_i$, regardless of the instantaneous PV output $p^g_i(\tau, t)$ \cite{kekatos2015stochastic}. As such, $q_{r,i}^{g}(\tau,t)$ generated by inverter $i$ is constrained as 
\begin{equation}
|q_{r,i}^{g}(\tau,t)|\leq \bar q^g_i:= \sqrt{(\bar s_i)^2-(\bar p^g_i)^2}, \quad \forall i \in \mathcal N_r ,t.
\end{equation}


\subsection{Two-timescale voltage regulation formulation}
Given two-timescale load consumption and generation that we model as Markovian processes \cite{carta2009review}, the task of voltage regulation is to find the optimal reactive power support per slot by configuring capacitors in every interval and adjusting inverter outputs in every slot, such that the \emph{long-term} average voltage deviation is minimized. As voltage magnitudes $\pmb{v}(\tau,t)$ depend solely on the control variables $\pmb{q}^g(\tau,t)$, they are expressed as implicit functions of $\pmb{q}^g(\tau,t)$, yielding $\pmb{v}_{\tau,t}(\pmb{q}^g(\tau,t))$, whose actual function forms for postulated grid models will be given Section \ref{sec:fast}. The novel two-timescale voltage control scheme entails solving the following stochastic optimization problem
\begin{subequations}\label{eq:twotimescale}
	\begin{align}
	\underset{\{\pmb{q}_r^g(\tau,t)\}\atop
		\left\{ \pmb{y}(\tau)\in \{0,1\}^{N_a}  \right \}}{\minimize}
	~~&{\mathbb{E}}\! \left[\sum_{\tau=1}^\infty \sum_{t=1}^{N_T}\gamma^\tau
	\left\|\pmb v_{\tau,t}(\pmb q^g(\tau,t))	- v_0\pmb{1}\right\|^2\right]\\
	{\rm subject\;to}\quad \;&q^g_i(\tau,t) =\hat y_{k_i}(\tau) q_{a,k_i}^{g}, 
	~\,\quad 
	\forall i\in\mathcal N_a, \tau, t\label{eq:const1} \\
	&q^g_i(\tau,t) = q_{r,i}^g(\tau,t),~\,\quad \quad
	\forall i\in \mathcal{N}_r, \tau, t\\
	&|q_{r,i}^{g}(\tau,t)|\leq \bar q^g_i,\qquad \qquad \forall i \in \mathcal N_r ,   \tau,t \label{eq:trad3} 
	\end{align}
\end{subequations}
for some discount factor $\gamma \in (0,1)$, where the expectation is taken over the joint distribution of $(\pmb{p}^c(\tau,t),\pmb{q}^c(\tau,t),\pmb{p}^g(\tau,t))$ across all intervals and slots. Clearly, the optimization problem \eqref{eq:twotimescale} involves infinitely many
variables $ \{\pmb{q}^g_r(\tau,t)\}$ and $\{\hat {\pmb{y}}(\tau) \}$, which are coupled across time via the cost function and the constraint \eqref{eq:const1}. 
Moreover, discrete variables $\hat{\pmb{y}}(\tau)\in\{ 0,1\}^{N_a} $ render problem \eqref{eq:twotimescale} nonconvex and generally \emph{NP-hard}. Last but not least, it is a multi-stage optimization, whose decisions are not all made at the same stage, and must also account for the power variability during real-time operation. In words, tackling \eqref{eq:twotimescale} exactly is challenging. 

Instead, our goal is to design algorithms that sequentially observe predictions $\{(\pmb{p}^c(\tau,t),\pmb{q}^c(\tau,t)),\pmb{q}^g(\tau,t)\}$, and solve near optimally problem \eqref{eq:twotimescale}. The assumption is that,  although no distributional knowledge of those stochastic processes involved is given, their realizations can be made available in real time, by means of e.g., accurate forecasting methods \cite{tsp2019psse}. In this sense, the physics governing the electric power system will be utilized together with data to solve \eqref{eq:twotimescale} in real time. Specifically, on the slow timescale, say at the end of each interval $\tau-1$, the optimal on-off capacitor decisions $\pmb y(\tau)$ will be set through a DRL algorithm that can learn from the predictions collected within the current interval $\tau-1$; while, on the fast timescale, namely at the beginning of each slot $t$ within interval $\tau$, our two-stage control scheme will compute the optimal setpoints for inverters, by minimizing the instantaneous bus voltage deviations while respecting physical constraints, given the current on-off commitment of capacitor units $\hat{\pmb y}(\tau)$ found at the very end of interval $(\tau -1)$. These two timescales are detailed in Sections~\ref{sec:fast} and \ref{sec:slow}, respectively.

\section{Fast-timescale Optimization of Inverters}\label{sec:fast}
As alluded earlier, the actual forms of $\pmb{v}_{\tau,t}(\pmb{q}^g(\tau,t))$ will be specified in this section, relying on the exact AC model or a linearized approximant of it.
Leveraging convex relaxation to deal with the nonconvexity, the considered AC model yields a second-order cone program (SOCP), whereas the linearized one leads to a linearly constrained quadratic program. In contrast, the latter offers an approximate yet computationally more affordable alternative to the former. Selecting between these two models relies on affordable computational capabilities. 

\subsection{Branch flow model}\label{subsec:branch}

\begin{figure}
	\centering
	\includegraphics[width =0.45 \textwidth]{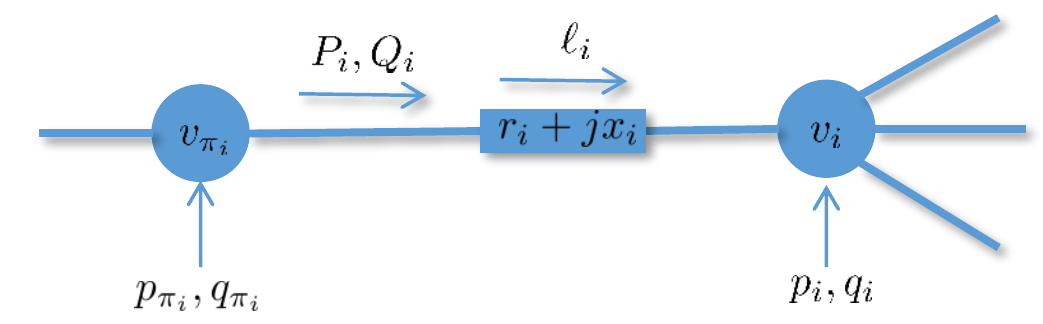}
	\caption{Bus $i$ is connected to its unique parent $\pi_i$ via line $i$.}
	\label{fig:lineardiagram}
\end{figure}

Due to the radial structure of distribution grids, every non-root bus $i\in\mathcal{N}$ 
has a unique parent bus termed $\pi_i$. The two are joined through the $i$-th distribution line represented by $(\pi_i,i) \in\mathcal{L}$ having impedance $r_i+jx_i$. Let $P_i(\tau, t)+jQ_i(\tau, t)$ stand for the complex power flowing from buses $\pi_i$ to $i$ seen at the `front' end at time slot $t$ of interval $\tau$, as depicted in Fig.~\ref{fig:lineardiagram}. Throughout this section, the interval index $\tau$ will be dropped when it is clear from the context.  

With further $\ell_i$ denoting the squared current magnitude on line $i\in \mathcal L$, the celebrated \emph{branch flow model} is described by the following equations for all buses $ i \in \mathcal N$, and for all $ t$ within every interval $\tau$ \cite{baran1989optimal,low2014convex} 
\begin{subequations}\label{eq:nonlinear1}
	\begin{align}
	p_i(t)&=\sum_{j\in\chi_i}P_j(t)  -( P_i(t) -r_i \ell_i(t)) \label{eq:nonp}\\
	q_i(t)&=\sum_{j\in\chi_i}Q_j(t)  - (Q_i(t)-x_i \ell_i(t))\label{eq:nonq}\\
	v_i(t)&={v_{\pi_i}(t)}\!- 2(r_iP_i(t)\!+x_iQ_i(t))\!+(r_i^2\!+x_i^2)\ell_i(t)  \label{eq:nonv}\\
	\ell_i(t)&=\frac{P^2_i(t)+Q^2_i(t)}{v_{\pi_i}(t)}\label{eq:ml}
	\end{align}
\end{subequations}
where we have ignored the dependence on $\tau$ for brevity, and $\chi_i$ denotes the set of all children buses for bus~$i$.

Clearly, the set of equations in \eqref{eq:ml} is quadratic in ${P}_i(t)$ and ${Q}_i(t)$, yielding a nonconvex set. To address this challenge, consider relaxing the equalities \eqref{eq:ml} into inequalities (a.k.a. hyperbolic relaxation, see e.g., \cite{farivar2011inverter}) 
\begin{align}\label{eq:inequality}
P^2_i(t)+Q^2_i(t)\le {v_{\pi_i}(t)}\ell_i(t), \quad \forall i \in \mathcal N,  t
\end{align}
which can be equivalently rewritten as the following second-order cone constraints
\begin{align}\label{eq:soc}
\!\left\|\begin{array}{c}
2P_i(t)\\
2Q_i(t)\\
\ell_i(t)-v_{\pi_i}(t)\end{array}\right\|\le {v_{\pi_i}(t)}+\ell_i(t), \quad \forall i \in \mathcal N.
\end{align}
Equations \eqref{eq:nonp}-\eqref{eq:nonv} and \eqref{eq:soc} now define a convex feasible set. 
The procedure of leveraging this relaxed set (instead of the nonconvex one) is known as SOCP relaxation \cite{low2014convex}. Interestingly, it has been shown that under certain conditions, SOCP relaxation is exact in the sense that the set of inequalties \eqref{eq:soc} holds with equalities at the optimum \cite{gan2015exact}. 

Given the capacitor configuration $\hat{\pmb y}(\tau)$ found at the end of the last interval $\tau-1$, under the aforementioned relaxed grid model, the voltage regulation on the fast timescale based on the exact AC model can be described as follows
\begin{subequations}\label{eq:nonlinear}
	\begin{align}
	\underset{\pmb{v}(t),\pmb{q}^g_r(t),
		\pmb{P}(t),\pmb{Q}(t)}{\minimize}~&~ \|\pmb v(t)
	- v_0\pmb{1}\|^2\\
	{\rm subject\;to}~~~\,
	~&~ \eqref{eq:nonp}-\eqref{eq:ml}\notag\\
	~&~q^g_i(t) = \hat y_{k_i}(\tau) q_{a,k_i}^{g},
	~\,\,\forall i\in\mathcal N_a \label{eq:trad11}\\
	~&~q^g_i(t) = q_{r,i}^g(t),~\,\quad \,~~~
	\forall i\in \mathcal{N}_r \label{eq:trad12}\\
	~&~|q_{r,i}^{g}(t)|\leq \bar q^g_i,\quad\quad\,\,\,\,\,\,\,\, \forall i \in \mathcal N_r  \label{eq:trad13} 
	\end{align}
\end{subequations}
which is readily a convex SOCP and can be efficiently solved by off-the-shelf convex programming toolboxes. 
The optimal setpoints of smart inverters for the exact AC model are found as the $\bm{q}_r^g$-minimizer of~$\eqref{eq:nonlinear}$.

However, solving SOCPs could be computationally demanding when dealing with relatively large-scale distribution grids, say of several hundred buses. Trading off modeling accuracy for computational efficiency, our next instantiation of the fast-timescale voltage control relies on an approximate grid model.

\subsection{Linearized power flow model}\label{sec:fastlinear}
As line current magnitudes $\{ \ell_i\}$ are relatively small compared to line flows, the last term in \eqref{eq:nonp}-\eqref{eq:nonv} can be ignored yielding the next set of linear equations for all $i,t$ \cite{baran1989network}
\begin{subequations}\label{eq:linearpowerflow}
	\begin{align}
	~&p_i(t)=\sum_{j\in\chi_i}P_j(t)- P_i(t)\label{eq:linp}\\
	~&q_i(t)=\sum_{j\in\chi_i}Q_j(t)  - Q_i(t)\label{eq:linq}\\
	~&v_i(t) =v_{\pi_i}(t)- 2(r_i P_i(t)+x_i Q_i(t))\label{eq:linv}
	\end{align}
\end{subequations}
which is known as the linearized distribution flow model. 
In this fashion, all squared voltage magnitudes $\pmb{v}(t)$ can be  expressed as linear functions of $\pmb q^g( t)$. 

Adopting the approximate model \eqref{eq:linearpowerflow}, the optimal setpoints of inverters can be found by solving the following optimization problem per slot $ t $ in interval $\tau$, provided $\hat {\pmb y} (\tau)$ is available from the last interval on the slow timescale
\begin{subequations}\label{eq:linear}
	\begin{align}
	\underset{\pmb{v}(t),\pmb{q}^g_r(t),
		\pmb{P}(t),\pmb{Q}(t)}{\minimize} ~&~ \|\pmb v(t)
	- v_0\pmb{1}\|^2\\
	{\rm subject~to}~~~~~&~\eqref{eq:linp}-\eqref{eq:linv} \notag\\
	~&~q^g_i(t) = \hat y_{k_i}(\tau) q_{a,k_i}^{g},	~\,\,\forall i\in\mathcal N_a \label{eq:trad21}\\
	~&~q^g_i(t) = q_{r,i}^g(t),~~~\,\quad \,\,\,\forall i\in \mathcal{N}_r\\
	~&~|q_{r,i}^{g}(t)|\leq \bar q^g_i,\quad\,\,\,\,\,\,\,\,\,~\,~ \forall i \in \mathcal N_r . \label{eq:trad23} 
	\end{align}
\end{subequations}

As all constraints are linear and the cost is quadratic, \eqref{eq:linear} constitutes a standard convex quadratic program. As such, it can be solved efficiently by e.g.,  primal-dual algorithms, or off-the-shelf convex programming solvers, whose implementation details are skipped due to space limitations. 

\section{Slow-timescale Capacitor Reconfiguration}\label{sec:slow}

Here we deal with reconfiguration of shunt capacitors on the slow timescale. This amounts to  determining their on-off status for the ensuing interval. Past approaches to solving the resultant integer-valued optimization were heuristic, or, relied on semidefinite programming relaxation. They do not guarantee optimality, while they also incur high computational and storage complexities. We take a different route by drawing from advances in artificial intelligence, to develop data-driven solutions that could near optimally learn, track, as well as adapt to unknown generation and consumption dynamics.  
%
%

\subsection{A data-driven solution} \label{subsec:datadriven}

\begin{figure}
	\centering
	\includegraphics[width =0.5 \textwidth]{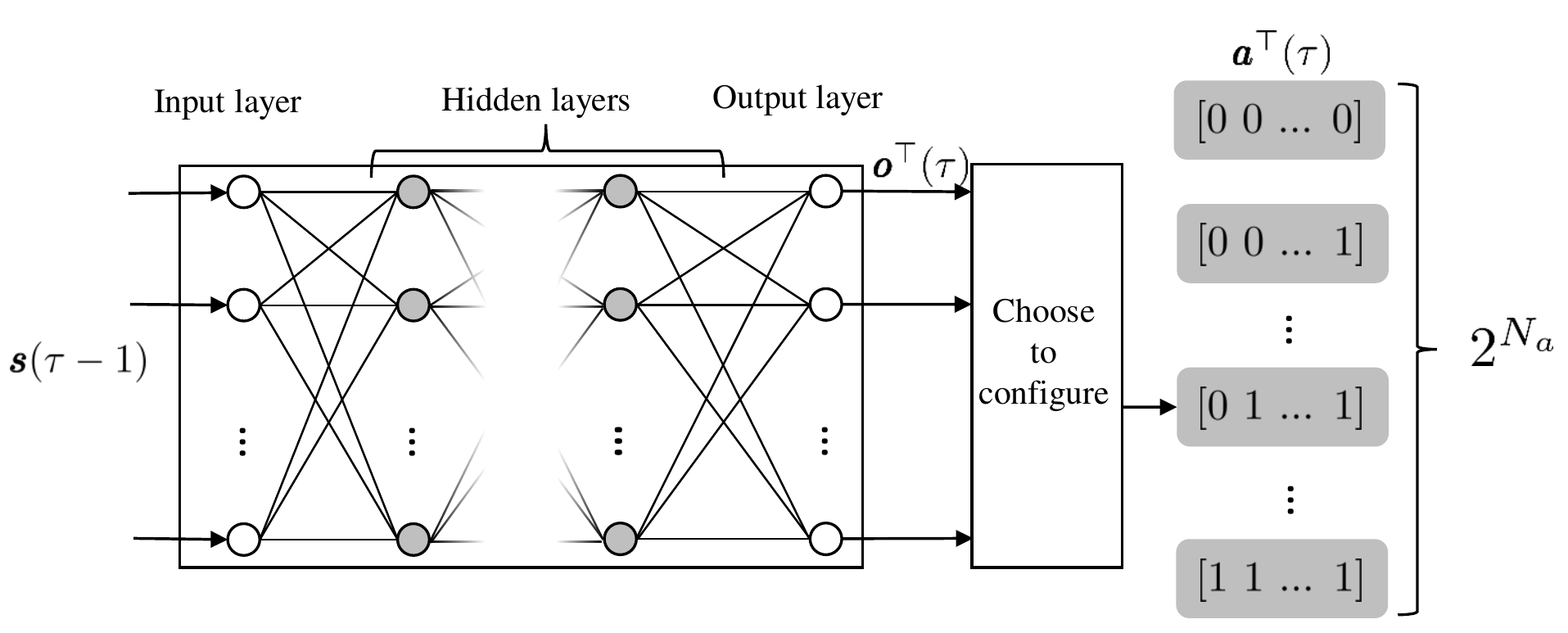}
	\caption{Deep $Q$-network}
	\label{fig:deepneuralnetwork}
\end{figure}

Clearly from \eqref{eq:trad11}--\eqref{eq:trad21}, the capacitor decisions  $ \hat {\pmb y}(\tau)$ made at the end of interval $\tau-1$ (slow-timescale learning) influence inverters' setpoints during the entire interval $\tau$ (fast-timescale optimization).  The other way around, inverters' regulation on voltages influences the capacitor commitment for the next interval. This two-way between the capacitor configuration and the optimal setpoints of inverters motivates our RL formulation.
Dealing with learning 
policy functions in an environment with action-dependent dynamically evolving states and costs, 
RL seeks a policy function (of states) to draw actions from, in order to minimize the average cumulative cost~\cite{RLbook}.


Modeling load demand and renewable generation as Markovian processes, the optimal configuration of capacitors can be formulated as an MDP, which can be efficiently solved through RL algorithms.
An MDP is defined as a 5-tuple $(\mathcal{S},\mathcal{A}, \mathcal{P}, c, \gamma)$, where $\mathcal {S}$
is a set of states; $\mathcal {A}$ is a set of actions; $\mathcal {P}$ is a set of transition matrices;  $c:\mathcal{S}\times \mathcal{A}\mapsto {\mathbb{R}}$ is a cost function such that, for $\pmb s \in \mathcal{S}$ and $\pmb a \in \mathcal{A}$, $c=(c(\pmb s, \pmb a))_{\pmb s\in\mathcal{S},\pmb a\in\mathcal{A}}$ are the real-valued instantaneous costs after the system operator takes an action $\pmb a$ at state $\pmb s$; and $\gamma\in [0,1)$ is the discount factor.
These components are defined next before introducing our voltage regulation scheme.

\textit{Action space $\mathcal{A}$}. Each action corresponds to one possible 
on-off commitment of capacitors $1$ to $N_a$, giving rise to an action vector $\pmb a(\tau)=\pmb y(\tau)$ per interval $\tau$.
The set of binary action vectors constitutes the action space $\mathcal{A}$, whose cardinality is exponential in the number of capacitors, meaning $|\mathcal{A}|=2^{N_a}$.

\textit{State space $\mathcal{S}$}. This includes per interval $\tau$ the average active power at all buses except for the substation, along with the current capacitor configurations; that is, $\pmb s(\tau) :=[{ {\,\bar{\pmb p}} }^{\top}(\tau), { \hat{\pmb y}}^{\top}(\tau)]^{\top}$, which contains both continuous and discrete variables. Clearly, it holds that $\mathcal{S}\subseteq \mathbb R^N \times 2^{N_a}$. 

The action is decided according to the configuration policy $\pi$ that is a function of the most recent state $\pmb s(\tau-1)$, given as
\begin{align}
\pmb a(\tau) = \pi(\pmb s(\tau-1)).
\end{align}

\textit{Cost function $c$}. 
The cost on the slow timescale is 
\begin{align}\label{eq:rlcost}
c(\pmb s(\tau-1), \pmb a(\tau)) = 	
\sum_{t=1}^{N_T}\!\left\|\pmb v_{\tau,t}(\pmb q^g(\tau,t)) - v_0\pmb{1}\right\|^2.
\end{align}

\textit{Set of transition probability matrices $\mathcal P$.} While being at a state $\pmb s \in {\mathcal S}$ upon taking an action $\pmb a$, the system moves to a new state $\pmb s ' \in {\mathcal S}$ probabilistically. Let $P^{\pmb a}_{\pmb s \pmb s'}$ denote the transition probability matrix from state $\pmb s$ to the next state $\pmb s'$ under a given action $\pmb a$. Evidently, it holds that ${\mathcal P} :=\left\{P^{\pmb a}_{\pmb s \pmb s'}| \forall {\pmb a} \in {\mathcal A} \right\}$. 

\textit{Discount factor $\gamma$.} The discount factor $\gamma \in [0,1)$, trades off the current versus future costs. The smaller $\gamma$ is, the more weight the current cost has in the overall cost. 

Given the current state and action, the so-termed action-value function under the control policy $\pi$ is defined as
\begin{align} \label{eq:Qdef}
&	Q_{\pi} (\pmb s(\tau-1), \pmb a(\tau))  :=  \notag\\ &{\mathbb{E}}\! \left[\sum_{\tau' = \tau}^{\infty}  \gamma^{\tau' - \tau} c(\pmb s(\tau'-1), \pmb a(\tau'))\Big | \pi , \pmb s(\tau-1),  \pmb a(\tau)\right]
\end{align}
where the expectation ${\mathbb{E}}$ is taken with respect to all sources of randomness.  

To find the optimal capacitor configuration policy $\pi^{\ast}$, that minimizes the average voltage deviation in the long run, we resort to the Bellman optimality equations; see e.g., \cite{RLbook}. Solving those yields the action-value function under the optimal policy $\pi ^*$ on the fly, given by 
\begin{equation}
\label{eq:Qoptimal}
Q_{\pi^*} \! \left(\pmb s, \pmb a\right) = {\mathbb{E}} \!\left[ c(\pmb s, \pmb a) \right] + \gamma \sum_{\pmb s' \in \mathcal{S}} P_{\pmb s \pmb s'}^{\pmb a} \min_{\pmb a \in \mathcal{A}} {Q_{\pi^*}} (\pmb s', \pmb a').
\end{equation}
With $Q_{\pi^{\ast}}(\pmb s, \pmb a)$ obtained, the optimal capacitor configuration policy can be found as 
\begin{equation}
\label{eq:optimal} 
\pi^{\ast} (\pmb s) = \arg \min_{\pmb a}\, Q_{\pi^{\ast}} (\pmb s, \pmb a ).
\end{equation}

It is clear from \eqref{eq:Qoptimal} that if all transition probabilities $\{P^{\pmb a}_{\pmb s \pmb s'}\}$ were available, we can derive $Q_{\pi^{\ast}}(\pmb s, \pmb a)$, 
and subsequently the optimal policy $\pi^{\ast}$ from \eqref{eq:optimal}. Nonetheless, obtaining those transition probabilities is impractical in practical distribution systems. This calls for approaches that aim directly at $\pi^*$, without assuming any knowledge of $\{P^{\pmb a}_{\pmb s \pmb s'}\}$. 

One celebrated approach of this kind is Q-learning, which can learn $\pi^{\ast}$ by approximating $Q_{\pi^{\ast}}(\pmb s, \pmb a)$ `on-the-fly'~\cite[p. 107]{RLbook}. Due to its high-dimensional continuous state space $\mathcal{S}$ however, $Q$-learning is not applicable for the problem at hand.
This motivates function approximation based $Q$-learning schemes that can deal with continuous state domains.

\subsection{A deep reinforcement learning approach} 
\label{subsec:DRL}

DQN offers a NN function approximator of the $Q$-function, chosen to be e.g., a fully connected feed-forward NN, or a convolutional NN, depending on the application \cite{mnih2015human}. It takes as input the state vector, to generate at its output $Q$-values for all possible actions (one for each). As demonstrated in \cite{mnih2015human}, such a NN indeed enables learning the $Q$-values of \textit{all} state-action pairs, from just a few observations obtained by interacting with the environment. Hence, it effectively addresses the challenge brought by the `curse of dimensionality'~\cite{mnih2015human}. Inspired by this, we employ a feed-forward NN to approximate the $Q$-function in our setting. Specifically, our DNN consists of $L$ fully connected hidden layers with ReLU activation functions, depicted in Fig.~\ref{fig:deepneuralnetwork}. At the input layer, each neuron is fed with one entry of the state vector $\pmb s(\tau-1)$, which, after passing through $L$ ReLU layers, outputs a vector $\pmb o(\tau)\in\mathbb{R}^{2^{N_a}}$, whose elements predict
the $Q$-values for all possible actions (i.e., capacitor configurations).  
Since each output unit corresponds to a particular configuration of all $N_a$ capacitors, there is a total of $2^{N_a}$ neurons at the output layer. 
For ease of exposition, let us collect all weight parameters of this DQN into a vector $\pmb \theta$ which parameterizes the input-output relationship as $\pmb o(\tau) = Q_{\pi}(\pmb s(\tau-1), \pmb a(\tau); \pmb \theta)$
(c.f.~\eqref{eq:Qdef}). 
At the end of a given interval $\tau-1$, upon passing the state vector $\pmb s(\tau-1)$ through this DQN, the corresponding predicted $Q$-values $\pmb o(\tau)$ for all possible actions become available at the output. 
Based on these predicted values, the system operator selects the action having the smallest predicted $Q$-value to be in effect over the next interval.

Intuitively, the weights $\pmb \theta$ should be chosen such that the DQN outputs match well the actual $Q$-values with input any state vector. Toward this objective, the popular stochastic gradient descent (SGD) method is employed to update $\pmb \theta$ `on the fly' \cite{mnih2015human}. 
At the end of a given interval $\tau$, precisely when i)  the system operator has made decision $\pmb a (\tau)$, ii) the grid has completed the transition from the state $\pmb s(\tau-1)$ to a new state $\pmb s(\tau)$, and, (iii) the network has incurred and revealed cost $c(\pmb s(\tau-1), \pmb a(\tau))$, we perform a SGD update based on the current estimate $\pmb \theta_\tau$ to yield $\pmb \theta_{\tau +1}$. The so-termed temporal-difference learning \cite{RLbook} confirms that a sample approximation of the optimal cost-to-go from interval $\tau$ is given by $c(\pmb s(\tau-1), \pmb a(\tau) ) + \gamma \min \limits_{\pmb a' \in \mathcal{A}} Q_{\pi}(\pmb s(\tau), \pmb a'; \pmb \theta_{\tau})$, where $c(\pmb s(\tau-1), \pmb a(\tau) )$ is the instantaneous cost observed, and $\min \limits_{\pmb a '}Q_{\pi}(\pmb s(\tau), \pmb a'; \pmb \theta_{\tau})$ represents the smallest possible predicted cost-to-go from state $\pmb s(\tau)$, which can be computed through our DQN with weights $\pmb \theta_{\tau}$, and is discounted by factor $\gamma$. In words, the target value  $c(\pmb s(\tau-1), \pmb a(\tau) ) + \gamma \min \limits_{\pmb a' \in \mathcal{A}} Q_{\pi}(\pmb s(\tau), \pmb a'; \pmb \theta_{\tau})$ is readily available at the end of interval $\tau-1$. 
Adopting the $\ell_2$-norm error criterion, a meaningful approach to tuning the weights $\pmb \theta$ entails minimizing the following loss function  
\begin{align}
\label{eq:errorDef}
{\mathcal L} (\pmb \theta) &:= \Big[c(\pmb s(\tau-1), \pmb a(\tau) ) + \gamma \min \limits_{\pmb a'\in \mathcal{A}} Q_{\pi}(\pmb s (\tau), \pmb a'; \pmb \theta_{\tau}) \notag\\ &\quad~~\,  - Q_{\pi}(\pmb s(\tau-1), \pmb a(\tau); \pmb \theta)\Big]^2
&	\end{align}
for which the SGD update is given by
\begin{equation}
\label{eq:SGD}
\pmb \theta_{\tau+1} = \pmb \theta_\tau - \beta_\tau \nabla {\mathcal L}(\pmb \theta)|_{{\pmb \theta}_\tau}
\end{equation} 
where $\beta_\tau>0$ is a preselected learning rate, and $\nabla {\mathcal L}(\pmb \theta)$ denotes the (sub-)gradient.

\begin{algorithm}[t]
	\caption{Two-timescale voltage regulation scheme.}
	\label{Alg_a}
	\begin{algorithmic}[1]
		\State \textbf{Initialize:} 
		$\pmb \theta_0$ randomly; weight of the target network $\pmb \theta^{\rm Tar}_0=\pmb \theta_0$; replay buffer $\mathcal R$; and the initial state $\pmb s(0)$.	
		\For{$\tau=1,2,... $ } 
		\State {Take action $\pmb a(\tau)$ through exploration-exploitation
			\begin{equation}
			\label{eq:actcapacitor}\notag
			\hspace{+0.4 cm} \pmb a(\tau)=\!\left\{\!\!\begin{array}{ll}
			\!{\rm { random }~\;~\pmb a \in {\mathcal A}}\!\!\!\!\!\!&{{\rm w.p.} \;\; \epsilon_\tau}\\
			\!	{\arg \min}_{\pmb a'}~ Q (\pmb s(\tau-1), \pmb a'; \pmb \theta_{\tau} )\!\!\!\!& {\rm w.p.~}{1\!-\!\epsilon_\tau}
			\end{array}
			\right.
			\end{equation}
			\hspace{+0.37 cm} {  where $\epsilon_\tau    = {\rm max} \big \{ 1 - 0.1  \times \lfloor \tau / 50 \rfloor, \, 0\big\}.$}}
		\State Evaluate $\pmb c(\pmb s(\tau-1),\pmb a (\tau))$ using \eqref{eq:rlcost}.
		\For{$t=1,2,...,N_T$} 
		\State Compute $\pmb q^g(\tau,t)$ using \eqref{eq:nonlinear} or \eqref{eq:linear}.
		\EndFor
		\State Update $\pmb s(\tau)$.
		\State Save $(\pmb s(\!\tau-\!1),\pmb a (\tau),c(\pmb s(\tau-1),\pmb a (\tau)),\pmb s(\tau))$ into $\mathcal R(\tau)$.
		\State Randomly sample $M_\tau$ experiences from $\mathcal R(\tau)$.  
		\State Form the mini-batch loss ${\mathcal{L}^{\rm Tar}}(\pmb \theta_\tau;\mathcal M_\tau)$ using \eqref{eq:minilosstar2}. 
		\State Update $\pmb \theta_{\tau+1}$ using \eqref{eq:miniSGD}.
		\If  {mod$(\tau,B)=0$}
		\State Update the target network $\pmb \theta^{\rm Tar}_\tau=\pmb \theta_\tau$.
		\EndIf
		\EndFor
	\end{algorithmic}
\end{algorithm}

However, due to the compositional structure of DNNs, the update \eqref{eq:SGD} does not work well in practice. In fact, the resultant DQN oftentimes does not provide a stable result; see e.g.,~\cite{wu2016google}. To bypass these hurdles, several modifications have been introduced. In this work, we adopt the \textit{target network} and \textit{experience replay} \cite{mnih2015human}. 
To this aim, let us define an experience $e (\tau'):=(\pmb s(\tau'-1),\pmb a (\tau')),c(\pmb s(\tau'-1),\pmb a (\tau')),\pmb s(\tau'))$, to be a tuple of state, action, cost, and the next state.  Consider also having a replay buffer $\mathcal{R}(\tau)$ on-the-fly, which stores the most recent $R>0$ experiences visited by the agent. For instance, the replay buffer at any interval $\tau\ge R$ is $\mathcal{R}(\tau):=\{e(\tau-R+1), 
\ldots, e(\tau)\}$. Furthermore, as another effective remedy to stabilizing the DQN updates, we replicate the DQN to create a second DNN, commonly referred to as the \textit{target} network, whose weight parameters are concatenated in the vector $\pmb \theta^{\rm Tar}$. It is worth highlighting that this target network is not trained, but its parameters $\pmb \theta^{\rm Tar}$ are only periodically reset to estimates of $\pmb \theta$, say every $B$ training iterations of the DQN.  
Consider now the temporal-difference loss for some randomly drawn experience $e (\tau')$ from $\mathcal{R}(\tau)$ at interval $\tau$ 
\begin{align}
\label{eq:minilosstar}
&{\mathcal{L}^{\rm Tar}}(\pmb \theta_\tau; e(\tau')) :=\frac{1}{ 2} \Big[{ c(\pmb s(\tau'-1),\pmb a(\tau'))}\notag \\
&+ \gamma  \!\min_{\pmb a'}  Q^{\rm Tar} (\pmb s(\tau), \pmb a';\pmb \theta^{\rm Tar}_{\tau'}) - Q(\pmb s(\tau'-1),\pmb a(\tau'); \pmb \theta_{\tau})\Big]^2.
\end{align}
Upon taking expectation with respect to all sources of randomness generating this experience, we arrive at 
\begin{equation}
\label{eq:minilosstar}
{\mathcal{L}^{\rm Tar}}(\pmb \theta_\tau; \mathcal R(\tau))) :=  \mathbb{E}_{e(\tau')} \, \mathcal{L}^{\rm Tar}(\pmb \theta_\tau;e(\tau')).
\end{equation}
In practice however, the underlying transition probabilities are unknown, which challenges evaluating and hence minimizing ${\mathcal{L}^{\rm Tar}}(\pmb \theta_\tau; \mathcal R(\tau))) $ exactly. A commonly adopted alternative is to approximate the expected loss with an empirical loss over a few samples (that is, experiences here). 
To this end, we draw a mini-batch of $M_\tau$ experiences uniformly at random from the replay buffer $\mathcal{R}(\tau)$, whose indices are collected in the set $\mathcal{M}_\tau$, i.e., $\{ e(\tau')\}_{\tau' \in \mathcal{M}_\tau}\sim U(\mathcal R(\tau))$.
Upon computing for each of those sampled experiences an output using the target network with parameters $\pmb \theta_\tau^{\rm Tar}$, the empirical loss is 
\begin{align}
\label{eq:minilosstar2}
&{\mathcal{L}^{\rm Tar}}(\pmb \theta_{\tau};\mathcal {M}_\tau) := \frac{1}{ 2M_\tau}\!\sum_{\tau' \in \mathcal{M}_\tau} \Big[{ c(\pmb s(\tau'-1),\pmb a(\tau'))}\notag \\
&+\gamma\! \min_{\pmb a'}  Q^{\rm Tar} (\pmb s(\tau'), \pmb a';\pmb \theta^{\rm Tar}_{\tau})  - Q(\pmb s(\tau'-1), \pmb a(\tau'); \pmb \theta_{\tau})  \Big]^2.
\end{align}

In a nutshell, the weight parameter vector $\pmb \theta_{\tau}$ of the DQN is efficiently updated `on-the-fly' using SGD over the empirical loss ${\mathcal{L}^{\rm Tar}}(\pmb \theta_{\tau};\mathcal {M}_\tau)$, with iterates given by
\begin{equation}
\label{eq:miniSGD}
\pmb \theta_{\tau+1} = \pmb \theta_{\tau} - \beta_\tau \nabla {\mathcal{L}^{\rm Tar}}(\pmb \theta_{\tau};\mathcal {M}_\tau).
\end{equation}

Incorporating \textit{target network} and \textit{experience replay} remedies for stable DRL, our proposed two-timescale voltage regulation scheme is summarized in Alg.~\ref{Alg_a}. 

\begin{figure}[t]
	\centering
	\includegraphics[width =0.49\textwidth]{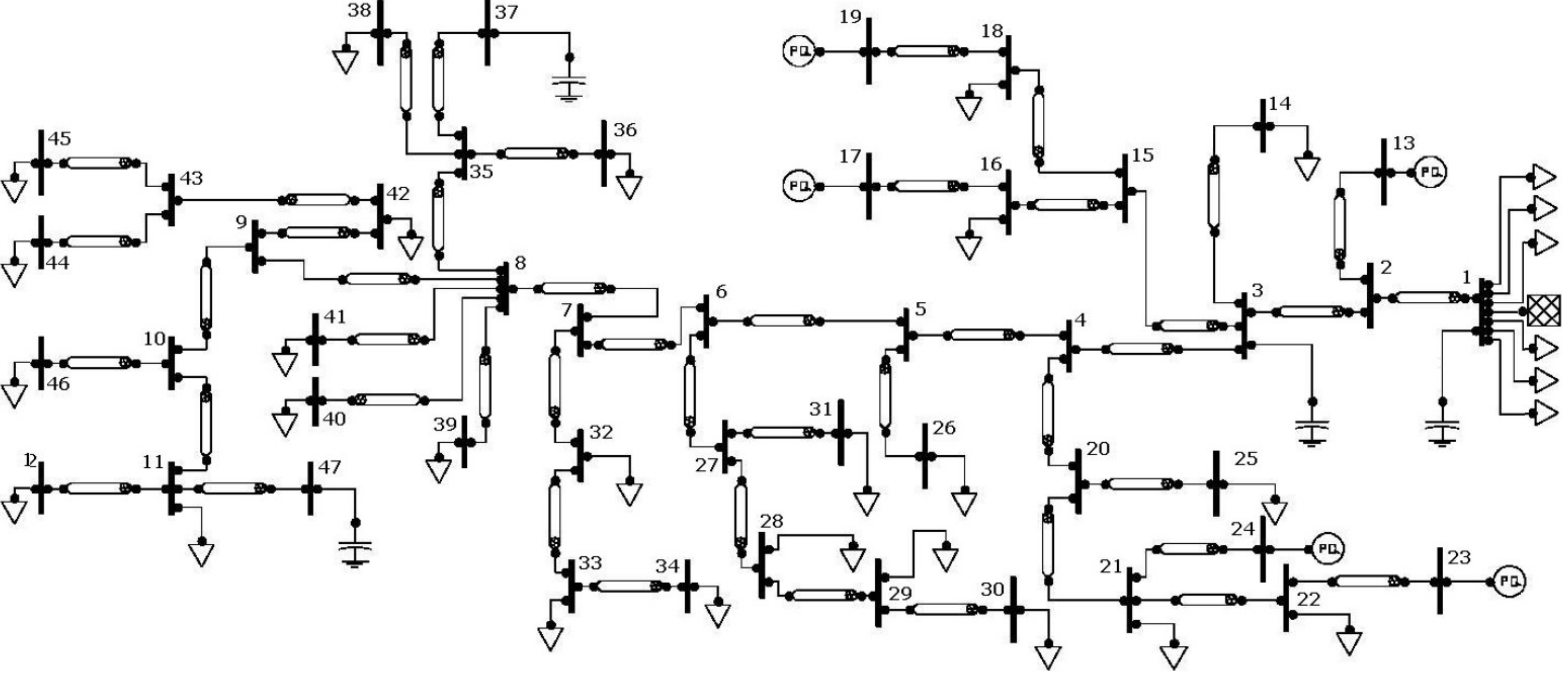}
	\caption{Schematic diagram of the $47$-bus industrial distribution feeder. Bus $1$ is the substation, and the $6$ loads connected to it model other feeders on this substation.
	}
	\label{fig:distributiongrid}
\end{figure}

\begin{figure}
	\centering
	\includegraphics[scale=.6]{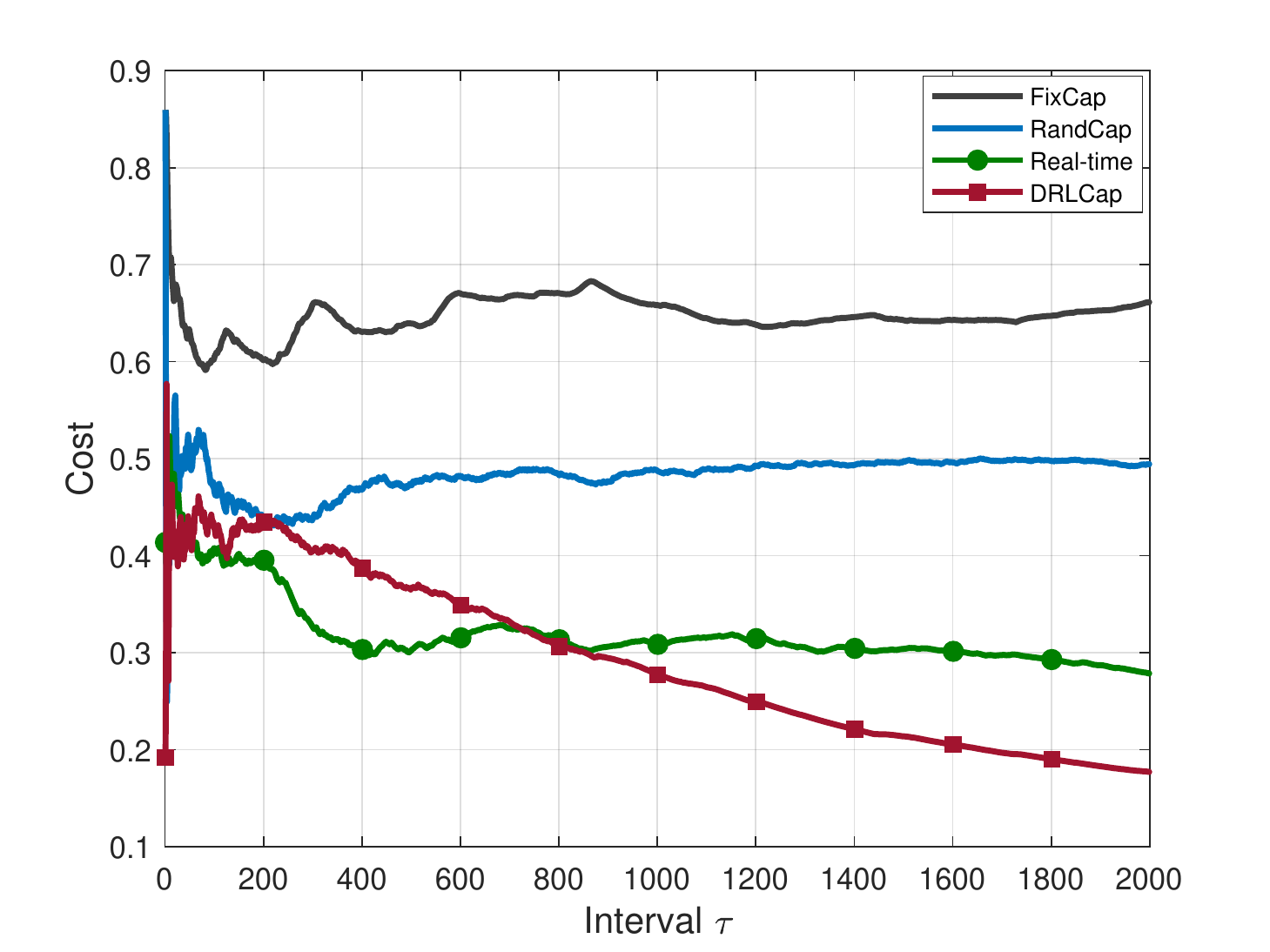}
	\caption{Time-averaged instantaneous costs incurred by the four voltage control schemes.}
	\label{fig:bus47cap3cost}
\end{figure}

\begin{figure}
	\centering
	\includegraphics[scale=.63]{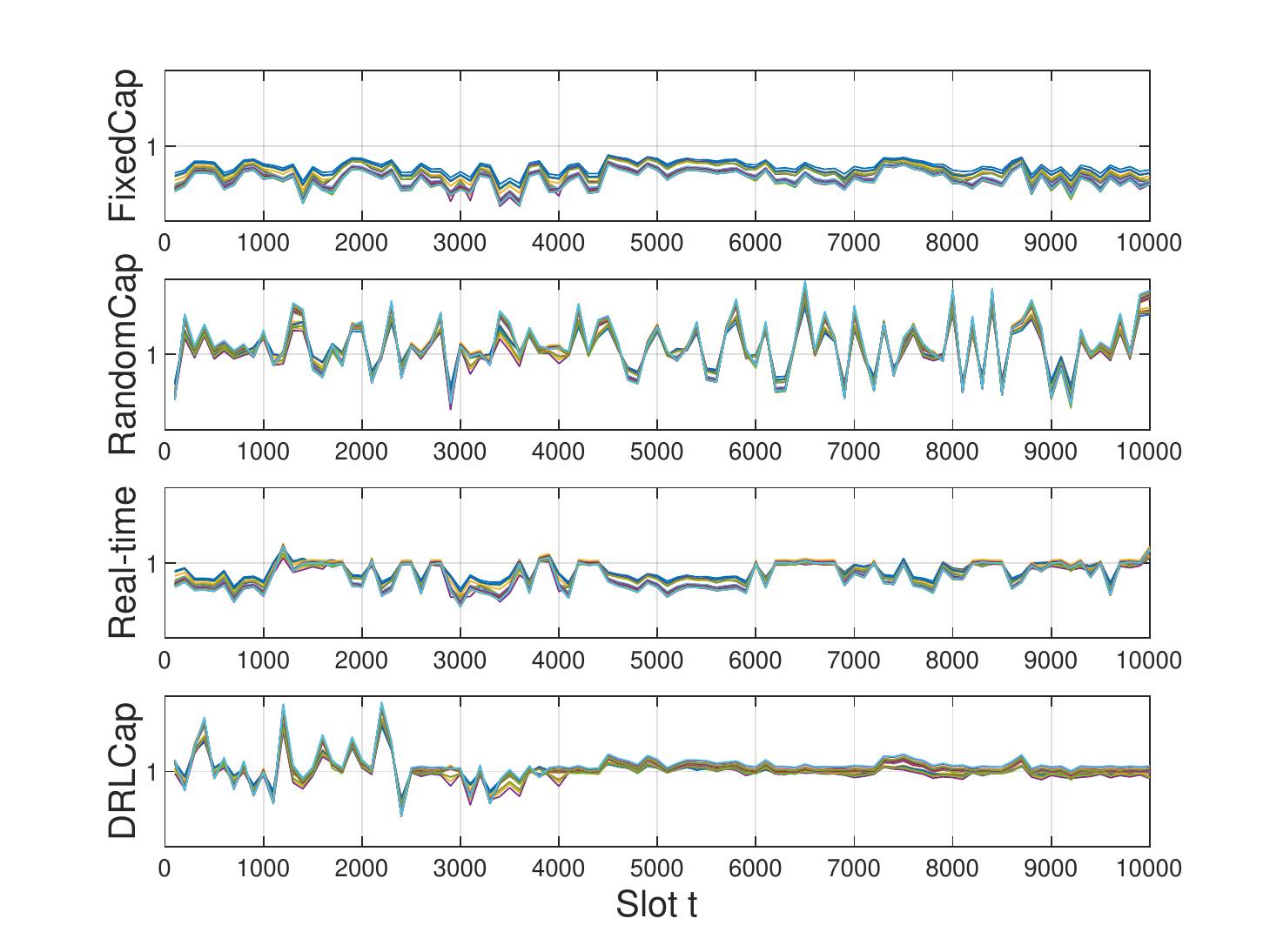}
	\caption{{ Voltage magnitude profiles obtained by the four voltage control schemes over the simulation period of $10,000$ slots.
	}}
	\label{fig:bus47cap3volt}
\end{figure}

\begin{figure}
	\centering
	\includegraphics[scale=.63]{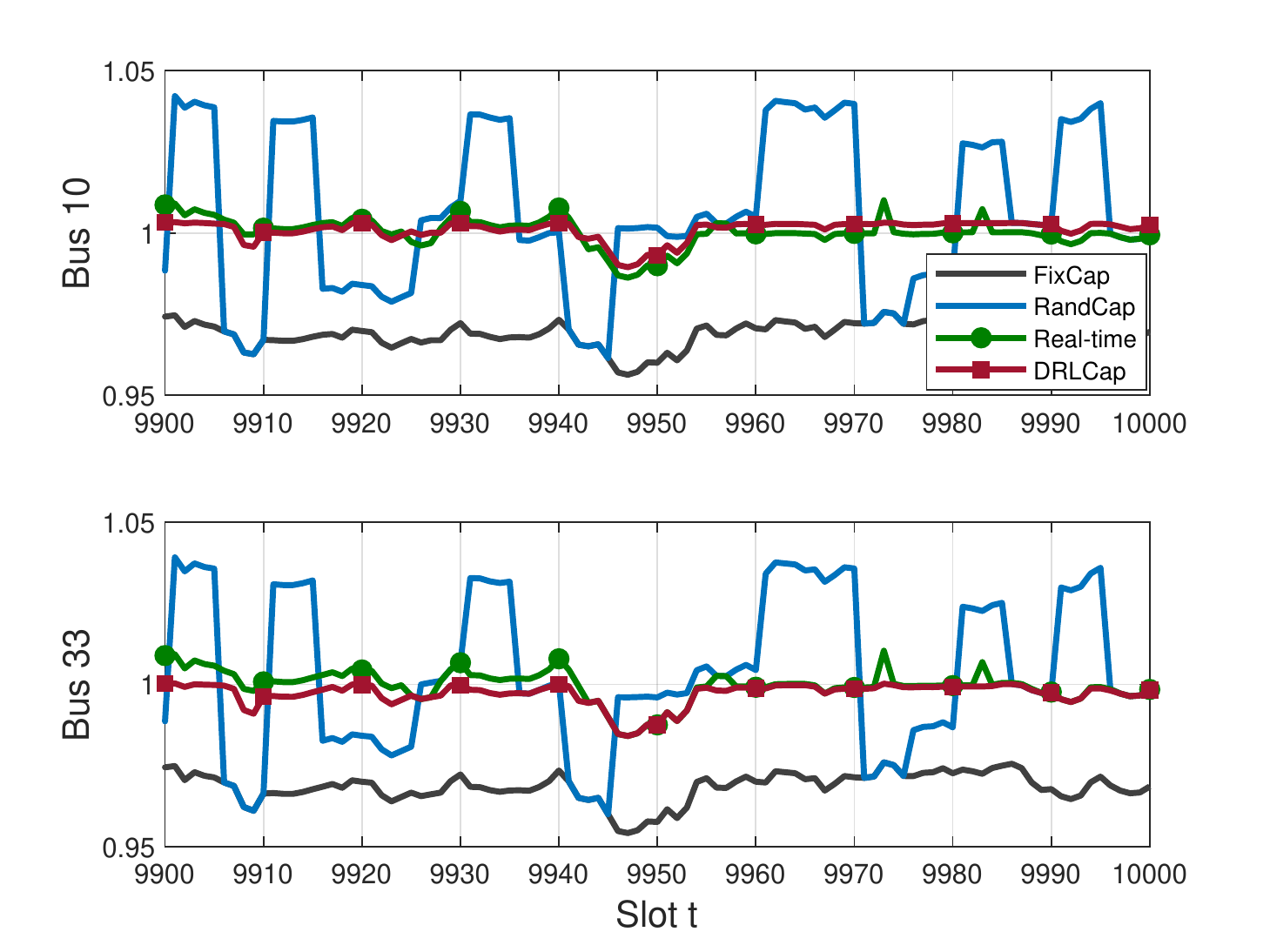}
	\caption{Voltage magnitude profiles obtained by the four voltage control schemes at buses $10$ and $33$ from slot $9,900$ to $10,000$.}
	\label{fig:bus47cap3volttwo}
\end{figure}

\begin{figure}
	\centering
	\includegraphics[scale=.63]{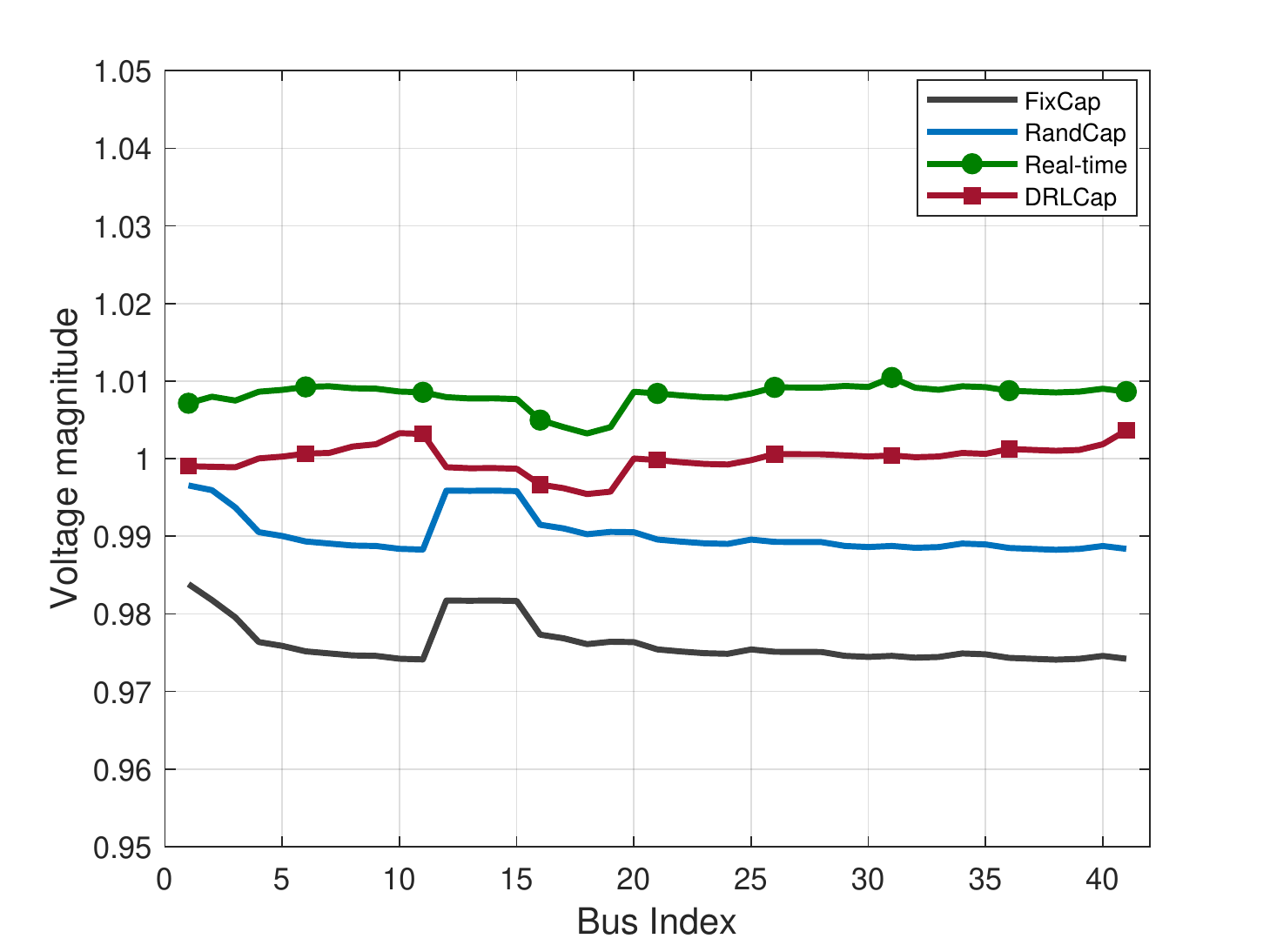}
	\caption{Voltage magnitude profiles at all buses at slot $9,900$ obtained by the four voltage control schemes. 
	}
	\label{fig:bus47cap3voltall}
\end{figure}

\section{Numerical Tests} \label{sec:test}

In this section, numerical tests on a real-world $47$-bus distribution feeder as well as the IEEE $123$-bus benchmark system are provided to showcase the performance of our proposed DRL-based voltage control scheme (cf. presented in Alg. \ref{Alg_a}). As has already been shown in previous works (e.g., \cite{kekatos2015stochastic,kekatos2016voltage,low2014convex}), the linearized distribution flow model approximates the exact AC model very well; hence, numerical results based on the linearized model were only reported here.  

The first experiment entails the Southern California Edison $47$-bus distribution feeder \cite{farivar2011inverter}, which is depicted in Fig.~\ref{fig:distributiongrid}. This feeder is integrated with four shunt capacitors
as well as five smart inverters. As the voltage magnitude $v_0$ of the substation bus is regulated to be a constant ($1$ in all our tests) through a voltage transformer, the capacitor at the substation was excluded from our control. Thus, a total of three shunt capacitors along with five smart inverters embedded with large PV plants were engaged in voltage regulation. 
The rest three capacitors are installed on buses $3$, $37$, and $47$, with capacities $120$, $180$, and $180$ kVar, respectively, while the five large PV plants are located on buses $2$, $16$, $18$, $21$, and $22$, with capacities $300$, $80$, $300$, $400$, and $200$ kW, respectively. 
To test our scheme in a realistic setting, 
real consumption as well as solar generation data were obtained from the Smart${^*}$ project collected on August $24, 2011$ \cite{barker2012smart}, which were first preprocessed by following the procedure described in our precursor work \cite{kekatos2015stochastic}.

In our tests, to match the availability of real data, each slot $t$ was set to a minute, and each interval $\tau$ was set to five minutes. A power factor of $0.8$ was assumed for all loads.
The DQN used here consists of three fully connected layers, which has $44$ and $12$ units in the first and second hidden layers, respectively. Although simple, it was found sufficient for the task at hand. ReLU activation functions ($\sigma(x)=\max(x,0)$) were employed in the hidden layers, and logistic sigmoid functions $s(x)=1/(1+e^{-x})$ were used at the output layer. 
To assess the performance of our proposed scheme, we have simulated three capacitor configuration policies as baselines, that include a fixed capacitor configuration (FixCap), a random capacitor configuration (RandCap), and an (impractical) `real-time' policy. Specifically, the FixCap uses a fixed capacitor configuration throughout, and the RandCap implements random actions to configure the capacitors on every slow time interval; both of which compute the inverter setpoints by solving  \eqref{eq:linear} per slot $t$. The impractical Real-time scheme however, optimizes over inverters and capacitors on a single-timescale, namely at every slot -- hence justifying its `real-time' characterization. To carry out this optimization task, first the binary constraints $y_{k_i}(t)\in \{0,1\}$ are relaxed to box ones $y_{k_i}(t)\in [0,1]$, the resulting convex program is solved using an off-the-shelf routine \cite{cvx}, which is followed by a standard rounding step to recover binary solutions for capacitor configurations \cite{tpd1989wu}.	 


In the first experiment, the DRL-based capacitor configuration (DRLCap) voltage control approach was examined. 
The replay buffer size was set to $R = 10$, the discount factor $\gamma = 0.99$, the mini-batch size $ M_\tau =10$, and the exploration-exploitation parameter $\epsilon_\tau    = {\rm max} \big \{ 1 - 0.1  \times \lfloor \tau / 50 \rfloor, \, 0\big\}.$
During training, the target network was updated every $B=5$ iterations.
The time-averaged instantaneous costs $$\frac{1}{\tau}\sum_{i=1}^\tau c(\pmb{s}(i-1),\pmb{a}(i))$$  
incurred by the four schemes over the first $1\le \tau\le 2,000$ intervals are plotted in Fig.~\ref{fig:bus47cap3cost}. Evidently, the proposed scheme attains a lower cost than FixCap, RandCap, and Real-time after a short period of learning and interacting with the environment. 
Even though the real-time scheme optimizes both capacitor configurations and inverter setpoints per slot $t$, its suboptimal performance in this case arises from the gap between the convexified problem and the original nonconvex counterpart.
{  Fig.~\ref{fig:bus47cap3volt} presents the voltage magnitude profiles for all buses regulated by the four schemes  sampled at every $100$ slots}. Again, after a short period ($\sim$\,$4,500$ slots) of training through interacting with the environment, our DRLCap voltage control scheme quickly learns a stable and (near-) optimal policy. 
In addition, voltage magnitude profiles regulated by FixCap, RandCap, Real-time, and DRLCap at buses $10$ and $33$ from slot $9,900$ to $10,000$ are shown in Fig.~\ref{fig:bus47cap3volttwo}, while the voltage magnitude profiles at
all buses at slot $9,900$ are presented in Fig.~\ref{fig:bus47cap3voltall}. 
Curves showcase the effectiveness of our DRLCap scheme in smoothing voltage fluctuations incurred due to large solar generation as well as heavy load demand.



\begin{figure}
	\centering
	\includegraphics[width =0.50 \textwidth]{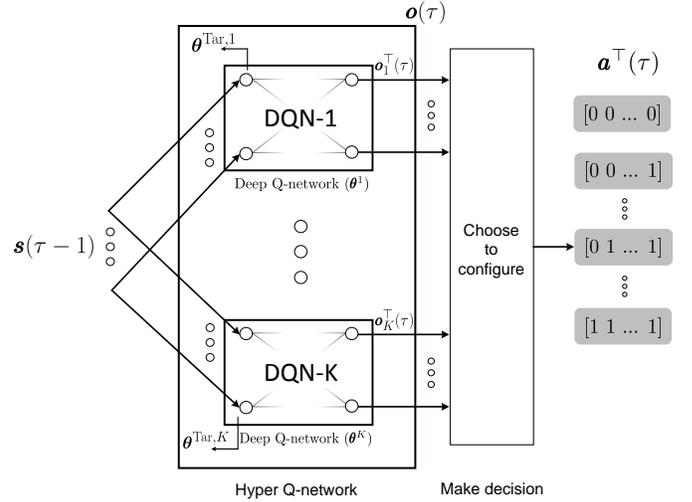}
	\caption{Hyper deep $Q$-network for capacitor configuration.}
	\label{fig:H_DQN}
\end{figure}

\begin{figure}[t]
	\centering
	\includegraphics[scale = .63 ]{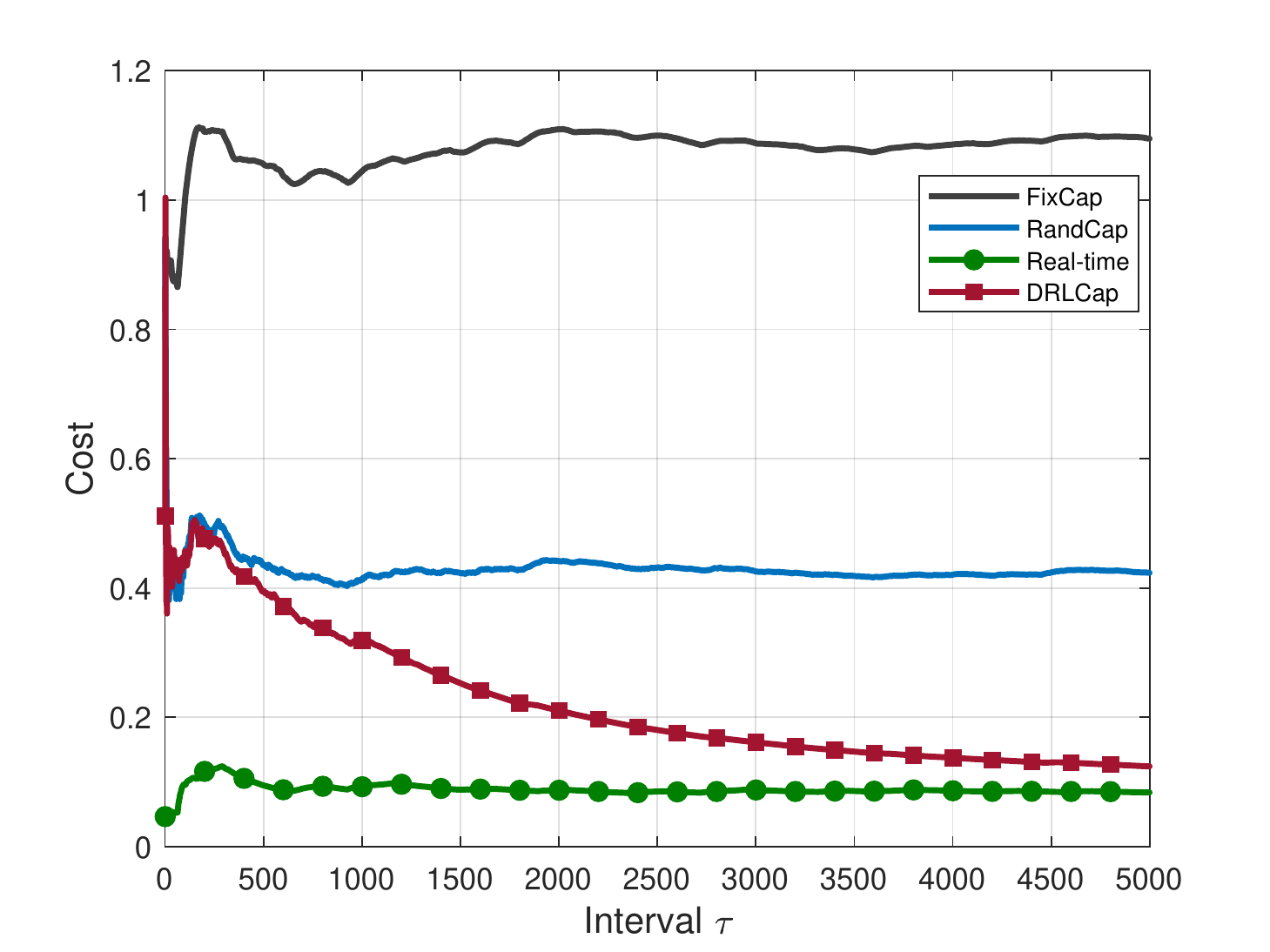}
	\caption{{  Time-averaged instantaneous costs incurred by the four approaches 
			on the IEEE $123$-bus feeder.}}
	\label{fig:bus123cap8cost}
\end{figure}

\begin{figure}
	\centering
	\includegraphics[scale=.63]{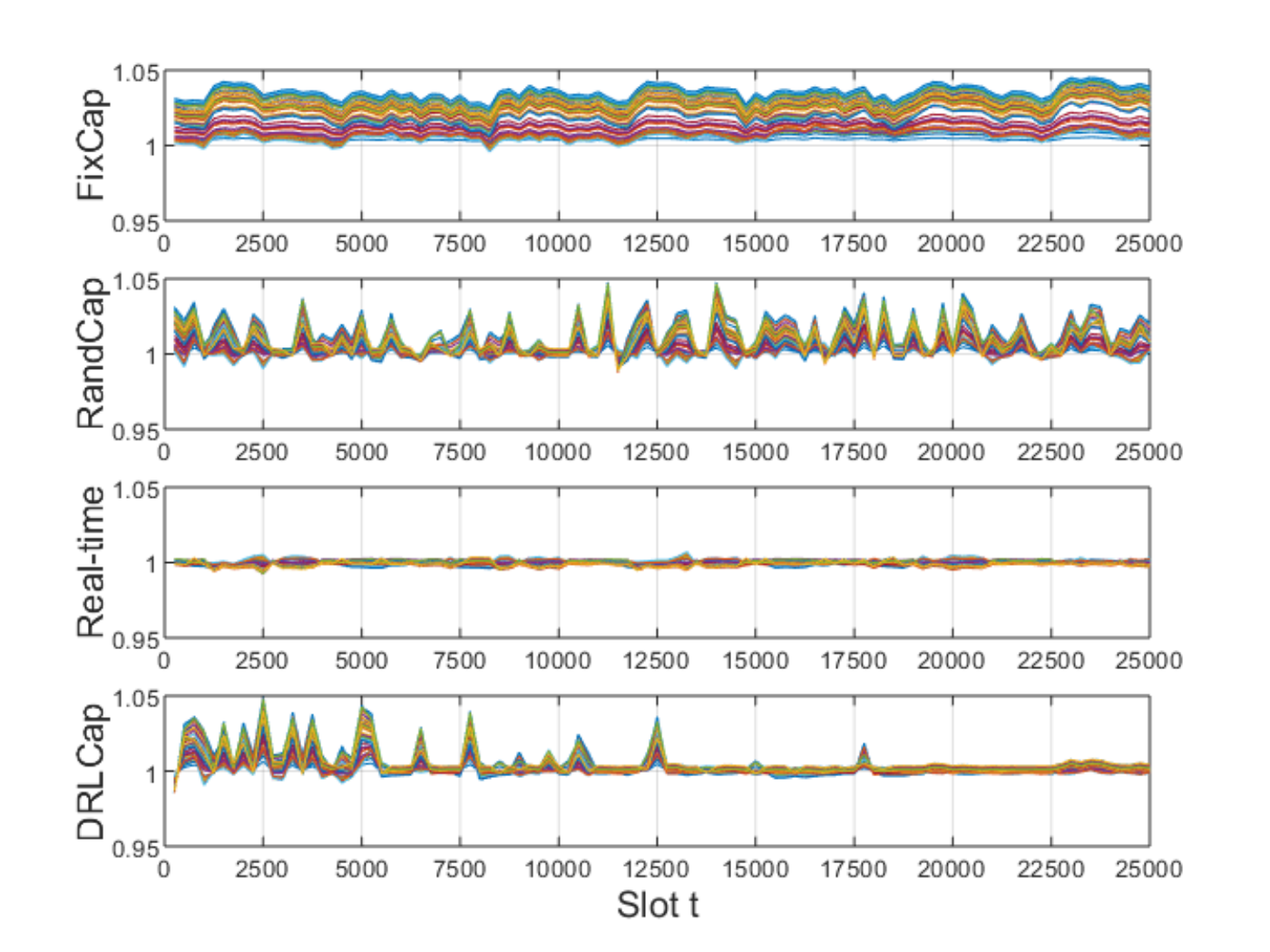}
	\caption{{ Voltage magnitude profiles at all buses over the simulation period of  $25,000$ slots on the IEEE $123$-bus feeder.}
	}
	\label{fig:bus123cap8volt}
\end{figure}

\begin{figure}[t]
	\centering
	\includegraphics[scale = .63]{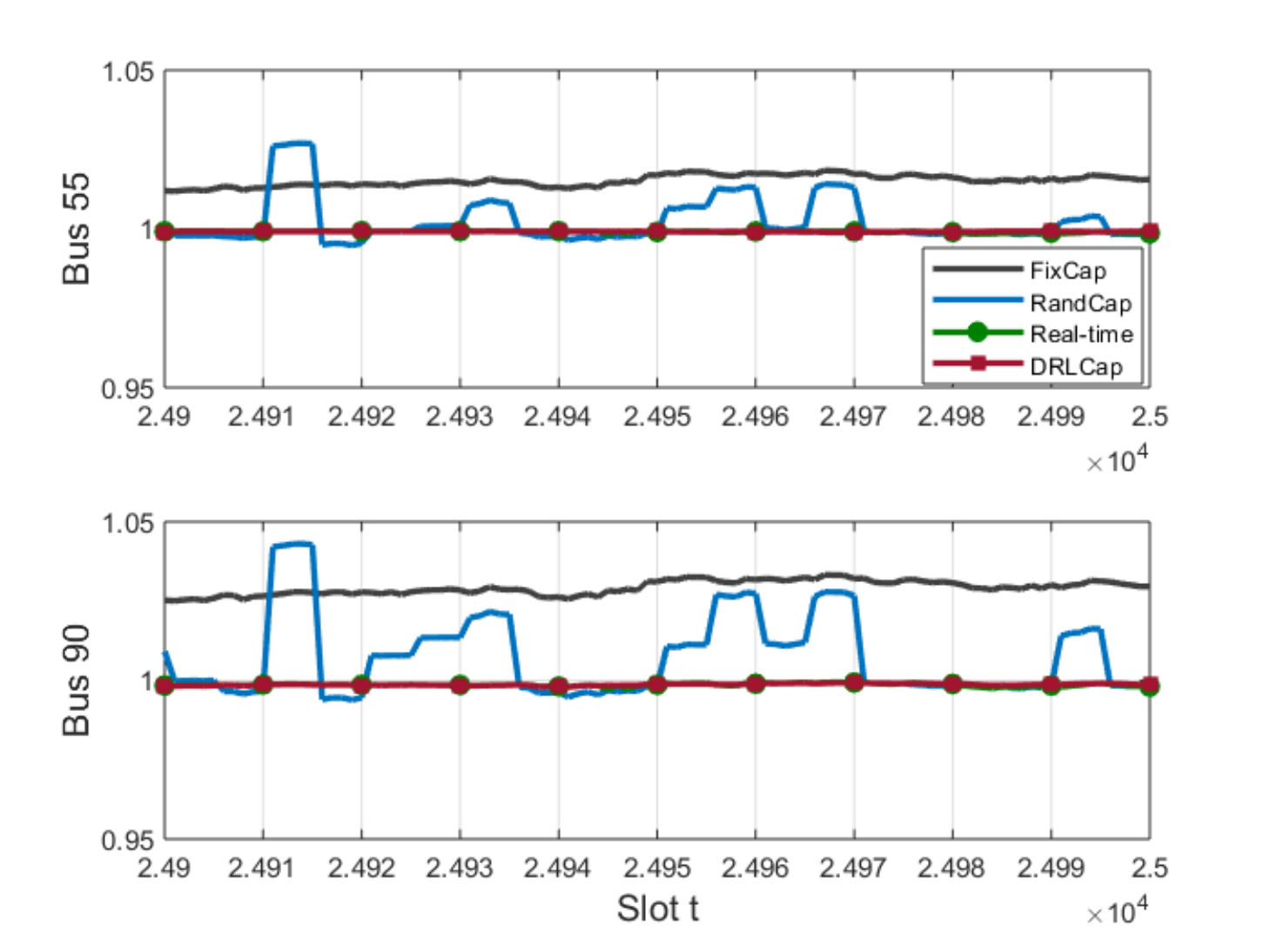}
	\caption{{ Voltage magnitude profiles at buses $55$ and  $90$ from slot $24,900$ to $25,000$ obtained by the four approaches 
			on the IEEE $123$-bus feeder.}}
	\label{fig:bus123cap8volttwo}
\end{figure}

\begin{figure}[t]
	\centering
	\includegraphics[scale = .63]{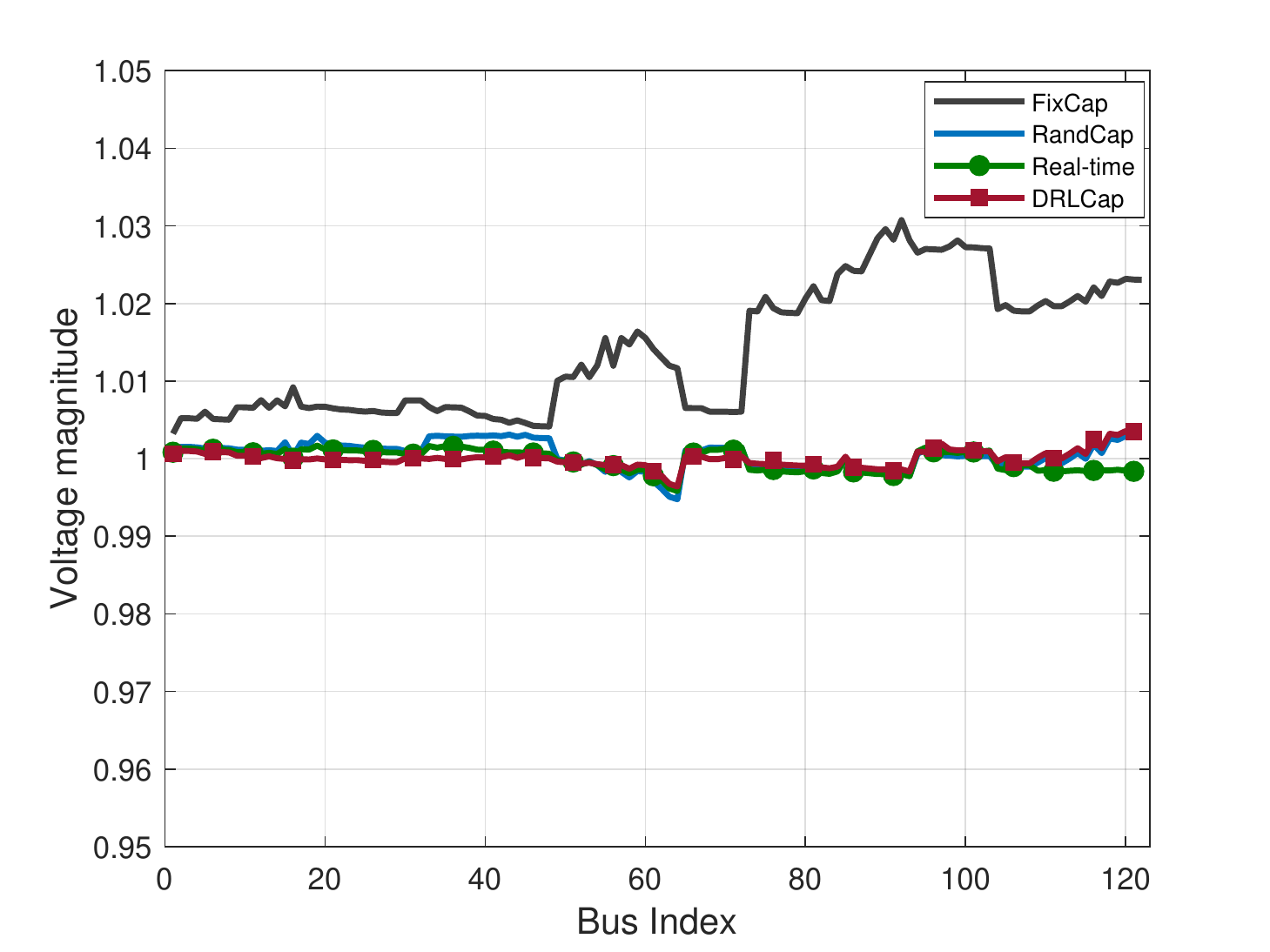}
	\caption{{ Voltage magnitude profiles at all buses on slot $24,900$ obtained by four approaches 
			on the IEEE $123$-bus feeder.}}
	\label{fig:bus123cap8voltall}
\end{figure}	

To deal with distribution systems having a moderately large number of capacitors, we further advocate a hyper deep $Q$-network implementation, that endows our DRL-based scheme with scalability. The idea here is to first split the total number $2^{N_a}$ of $Q$-value predictions $\pmb o(\tau)\in\mathbb{R}^{2^{N_a}}$ at the output layer into $K$ smaller groups, each of which is of the same size $2^{N_a}/K$ and is to be predicted by a small-size DQN. This evidently yields the representation $\pmb o(\tau) := [{\pmb o}_1^{\top}(\tau), \ldots, {\pmb o}_K^{\top}(\tau)]^\top \!\!$, where ${\pmb o}_k(\tau) \in {\mathbb R}^{2^{N_a}/K}\!\!$ for $k = 1, \ldots, K$. By running $K$ DQNs in parallel along with their corresponding target networks, each DQN-$k$ generates  predicted $Q$-values $\pmb o_k(\tau)$ for the subset of actions corresponding to $k$th group. Note that all DQNs are fed with the same state vector $\pmb s(\tau-1)$; see also Fig. \ref{fig:H_DQN} for an illustration. 

To examine the scalability and performance of this hyper $Q$-network implementation, additional tests using the IEEE $123$-bus test feeder with $9$ shunt capacitors were performed.
Again, the capacitor at bus $1$ was excluded from the control, rendering a total number of $2^8=256$ actions (capacitor configurations).
Renewable (PV) units are located on buses $47$, $49$, $63$, $73$, $104$, $108$, $113$, with capacities $100$, $16$, $70$, $20$, $20$, $30$, and $10$ k, respectively.   
The $8$ shunt capacitors are installed on buses $3$, $20$, $44$, $93$, $96$, $98$, $100$, and $114$, with capacities $50$, $80$, $100$, $100$, $100$, $100$, $100$, and $60$ kVar.
In this experiment, we used a total of $K = 64$ equal-sized DQNs to form the hyper $Q$-network, where each DQN implemented a fully connected $3$-layer feed-forward neural network, with ReLU activation functions in the hidden layers, and sigmoid functions at the output. 
The replay buffer size was set to $R = 50$, the batch size to $M_\tau=8$, and the target network updating period to $B=10$. 
{ The time-averaged instantaneous costs obtained over a simulation period of $5,000$ intervals 
	is plotted in Fig.~\ref{fig:bus123cap8cost}. 
	Moreover, voltage magnitude profiles of all buses over the simulation period of $25,000$ slots sampled at every $100$ slots under the four schemes are plotted in Fig.~\ref{fig:bus123cap8volt};
	voltage magnitude profiles at buses $55$ and $90$ from slot $24,900$ to $25,000$ are shown in Fig.~\ref{fig:bus123cap8volttwo}; and, voltage magnitude profiles at all buses on slot $24,900$ are depicted in  \ref{fig:bus123cap8voltall}. } 
Evidently, the hyper deep $Q$-network based DRL scheme smooths out the voltage fluctuations after a certain period ($\sim 7,000$ slots) of learning, while effectively handling the curse of dimensionality in the control (action) space. Evidently from Figs. \ref{fig:bus123cap8cost} and \ref{fig:bus123cap8voltall}, both the time-averaged immediate cost as well as the voltage profiles of DRLCap converge to those of the impractical `real-time' scheme (which jointly optimizes  inverter setpoints and capacitor configurations per slot). 


\section{Conclusions}
\label{sec:conc}
In this work, joint control of traditional utility-owned equipment and contemporary smart inverters for voltage regulation through reactive power provision was investigated. To account for the different response times of those assets, a two-timescale approach to minimizing bus voltage deviations from their nominal values was put forth, by combining physics- and data-driven stochastic optimization. Load consumption and active power generation dynamics were modeled as MDPs. On a fast timescale, the setpoints of smart inverters were found by minimizing the instantaneous bus voltage deviations, while on a slower timescale, the capacitor banks were configured to minimize the long-term expected voltage deviations using a deep reinforcement learning algorithm. The developed two-timescale voltage regulation scheme was found efficient and easy to implement in practice, through extensive numerical tests on real-world distribution systems using real solar and consumption data. This work also opens up several interesting directions for future research, including deep reinforcement learning for real-time optimal power flow as well as unit commitment. 

\bibliographystyle{IEEEtran}
\bibliography{power}

\end{document}